\documentclass[a4paper,10pt]{article}
\usepackage{epsfig}

\begin{document}

\title{Supergiant halos as an integral record of natural pionic radioactivity}
\author{D. B. Ion$^{1}$, Reveica Ion-Mihai$^{2}$, M. L. Ion$^{2}$ and \\
%EndAName
Adriana I. Sandru$^{1}$ \\
$^{1}$ \textit{National Institute for Physics and Nuclear Engineering,}\\
\textit{IFIN-HH}, Bucharest, Romania\\
$^{2}$ \textit{University of Bucharest, Department of Atomic and} \\
\textit{Nuclear Physics}, Bucharest, Romania}
\maketitle

\begin{abstract}
In this paper an unified interpretation of the supergiant halos
(SGH), discovered by Grady, Walker and Laemmlein, is discussed.
So, it is proved that SGH`s can be considered as integral records
of the nuclear pionic radioactivity.
\end{abstract}

The radioactive halos (see Ref. [1]) were first reported between 1880 and
1890 and their origin was a mystery until the discovery of radioactivity and
its power of coloration. The radioactive halos are spherical,
microscopic-sized discolourations in crystals. In cross sections on a
microscope slide, they appear as a series of tiny concentric rings, usually
surrounding a central core (see Fig. 1). This central core is (or at least
initially was) an radioactive inclusion in crystal. The $\alpha $
-particles, emitted from the radioactive inclusion during radioactive decay,
damage the mineral and discolor it, with most of the damage occurring where
the particle stop. How far this $\alpha $-particle travels depend on its
energy. Since all the $\alpha $-particles from a particular parent nucleus
have the same energy and the particles are fired in all directions, a
spherical shell of discolouration will produced, appearing circular in
cross-section. Radioactive uranium $^{238}$U from inclusions generates
multi-ringed halo (see Fig. 1) because its radioactive decay series:
\begin{equation}
\begin{tabular}{l}
$^{238}U\left( \alpha \right) \rightarrow ^{234}Th\left( \beta \right)
\rightarrow ^{234}Pa\left( \beta \right) \rightarrow ^{234}U\left( \alpha
\right) \rightarrow ^{230}Th\left( \alpha \right) \rightarrow$ \\
$^{226}Ra\left( \alpha \right) \rightarrow ^{222}Rn\left( \alpha
\right) \rightarrow ^{218}Po\left( \alpha \right) \rightarrow
^{214}Pb\left( \beta
\right) \rightarrow ^{214}Bi\left( \beta \right) \rightarrow$ \\
$^{214}Po\left( \alpha \right) \rightarrow ^{210}Pb\left( \beta
\right) \rightarrow ^{210}Bi\left( \beta \right) \rightarrow
^{210}Po\left( \alpha
\right) \rightarrow ^{206}Pb\left( stable\right) $%
\end{tabular}
\label{1}
\end{equation}

Hence, of the 15 parent nuclei in this $^{238}$U-decay chain, eight emit
alpha particles when they decay forming eight rings. However, due to
overlap, only five of eight rings of a $^{238}$U halo are normally visible.
If, instead of radioactive uranium $^{238}$U, the inclusion was composed of
an radioactive isotope along the decay chain, there would be only fewer
rings. So, omitting the first few isotopes in the decay series it is quite
simple to work out which isotope was originally in the inclusion by counting
the rings. Hence, $^{218}$Po forms three rings, $^{214}$Po form two rings
while $^{210}$Po forms only one.

Therefore, aside from their interest as attractive mineralogical oddities,
the halos are of great interest for the nuclear physics because they are an
integral record of radioactive nuclear decay in minerals. This integral
record is detailed enough to allow estimation of the energy involved in the
decay process and to identify the decaying nuclide through genetic decay
chain. This latter possibility is particularly exciting because there exist
certain classes of halos, such as [1]: dwarf halos, X-halos, giant halos and
the supergiant halos [3-4], which cannot be identified with the ring
structure of the known alpha-emitters. Hence, barring the possibility of a
non radioactive origin, these new variants of halos can be interpreted as
evidences for hitherto undiscovered alpha-radionuclide, as well as, signals
for the existence of new types of radioactivities.

Therefore we must underline that in addition to the Laemmlein discovery [4],
who found halos of rather diffuse boundaries with radii up to several
thousand micrometers surrounding Thorium-containing monazite inclusions in
quartz, Grady and Walker [3] found twenty-five extremely large halos called
supergiant halos (SGH).

The SGH have the following essential characteristic features:

1. The SGH radii are between 50$\mu m$ and 410$\mu m$ . They have large oval
to circular inclusions with radii of 20-52 $\mu m.$

2. SGH do not have sharp edges.

3. In thick sections ( ${>}30\mu m$ ) SGH are surrounded by
approximately circular brown regions having the same color in
unpolarized light as the normal halos and exhibiting the same
colors of pleochroism..

4. The coloration of SGH gradually fades as they move away from inclusion.

5. The SGH, with inclusions removed and etched with HF, show an abundance of
fossil fission tracks in immediate proximity to the inclusion.

6. Three-dimensional etching of the SGH revealed a complex pattern of
extended fossil fission tracks distribution throughout the halo.

7. The SGH inclusions were found to be alpha-active.

Hence, barring the possibility of a non radioactive origin, these new
variants of halos can be interpreted as evidences for hitherto undiscovered
alpha-radionuclide, as well as, signals for the existence of new types of
radioactivities.

In the papers [5] we discussed a completely new possibility for an unified
interpretation of the supergiant halos (SGH), namely, that SGH`s are the
integral records of the nuclear pionic radioactivity [5-7]. Indeed, the
nuclear pionic radioactivity of a parent nucleus (A,Z) can be considered as
an inclusive reaction of form:

\begin{equation}
(A,Z)\rightarrow \pi +X  \label{2}
\end{equation}

where X denotes any configuration of final particles (fragments,
light neutral and charged particles, etc.) which accompany
emission process. The inclusive NPIR (Nuclear PIonic
Radioactivity) is in fact a sum of all exclusive nuclear
reactions allowed by the conservation laws in which a pion can be
emitted by a nucleus from its ground state. The most important
exclusive reactions which give the essential contribution to the
inclusive NPIR (1) are the spontaneous pion emission accompanied
by two body fission:
\begin{equation}
_{Z}^{A}X\rightarrow \pi +_{Z_{1}}^{A_{1}}X+_{Z_{2}}^{A_{2}}X  \label{3}
\end{equation}

Charged pions $\pi ^{\pm }$ as well as neutral pions $\pi ^{0}$ can be
emitted during two body or many body fission of parent nucleus. Therefore,
we must show that a nuclear reaction (2) is able to produce the supergiant
halos discovered by Grady and Walker [3] and Laemmlein [4].

Now, we remember that in order for a ''discolored'' region in a transparent
mineral truly to be a radioactive halo (RH), it must satisfy the following rules:

R1: It must have an inclusion that is, or at one time was, radioactive. The
dimension of the pionic radiohalo (PIRH) inclusion must be sufficiently large in order to
satisfy the dose rule R2.

R2: The dimension of a radioactive halo is given by that particle emission
process with largest range with the condition that its dose satisfy the
coloration threshold condition.

R3: In the nuclear reactions (1) from the PIRH-inclusions, the $\pi ^{-}$-
yields should be about two orders of magnitude larger than the $\pi ^{+}$
yields (see Ref. [5])

A radioactive halo is called pionic radiohalo (PIRH) if its
inclusion (R1) contains, or at one time has contained, a parent pionic
radioactive (1) nuclide in such a concentration that produced a pionic dose
satisfying the coloration threshold condition (R2). The PIRH by definition
must be the integral record in time of the nuclear pionic radioactivity in
some minerals such as biotite, fluorite, cordierite, etc. According to
R3-rule (see Ref. [5]), we expect that, two types of PIRH are possible to
be observed in nature. These will be the following:

\begin{itemize}
\item PIRH(-), when only $\pi ^{-}$ satisfy the dose rule R2 [5];

\item SPIRH, which is defined just as a superposition of PIRH(-) and PIRH(+), when
both $\pi ^{+}$ and $\pi ^{-}$ satisfy the dose rule R2 .
\end{itemize}

  PIRH(-) and SPIRH have the same signatures excepting a big difference
(about 700-800 $\mu m$ in radius, in favor of SPIRH) in dimensions. This
difference is given by the fact that $\pi ^{+}-$ meson are mainly decaying
via $\pi ^{+}\rightarrow \mu ^{+}+\nu _{\mu }$, while the $\pi ^{-}$ are
stopped in mineral producing the fission of the mineral nuclei in two or
more fragments.

The PIRH signatures are as follows [5]:

\begin{itemize}
\item  PIRH is a pionic radiohalo of supergiant dimensions which possess a
large inclusion, sufficient to satisfy the rule R2. The supergiant PIRH
dimension is essentially determined by the high ranges of $\pi$-mesons as
well as of their product of decay $\mu $-mesons (see Fig. 2);

\item  The PIRH do not have sharps edges and their coloration gradually
fades as moves away from inclusion;

\item  The PIRH, must show an abundance of fossil fission tracks in the
immediate proximity to the inclusion. This signature is given just by the
fission fragments of the pionic emitters (see reaction (1)) from the
PIRH-inclusion.

\item  The PIRH inclusions, by definition, can be $\alpha $ -active. As a
consequence of the alpha-activity of the nuclei from the PIRH inclusions
they must be surrounded by approximately circular regions having the same
color as normal halos and exhibiting the same colors of pleochroism;

\item  Extended fossil fission tracks distribution through the PIRH must be
also present as a consequence of the fission of the mineral nuclei
produced by the stopping $\pi ^{-}$ mesons in mineral.
\end{itemize}

Next, comparing the PIRH signatures with the above essential characteristic
features (1-7) of the SGH, it is easy to see that the PIRH(-) can be
identified, with a surprising high accuracy, as being the SGH discovered by
Grady and Walker [3] in biotite.

We note that a special investigation of the discovery of the Laemmlein-like
[4] halos are necessary, since these SGH with dimensions around 1-1.5 mm can
be identified as being the SPIRH produced by emission of $\pi
^{+}$-mesons during fission of the parent nuclei from the halos inclusions.

\begin{center}
\epsfig{width = 70mm, file=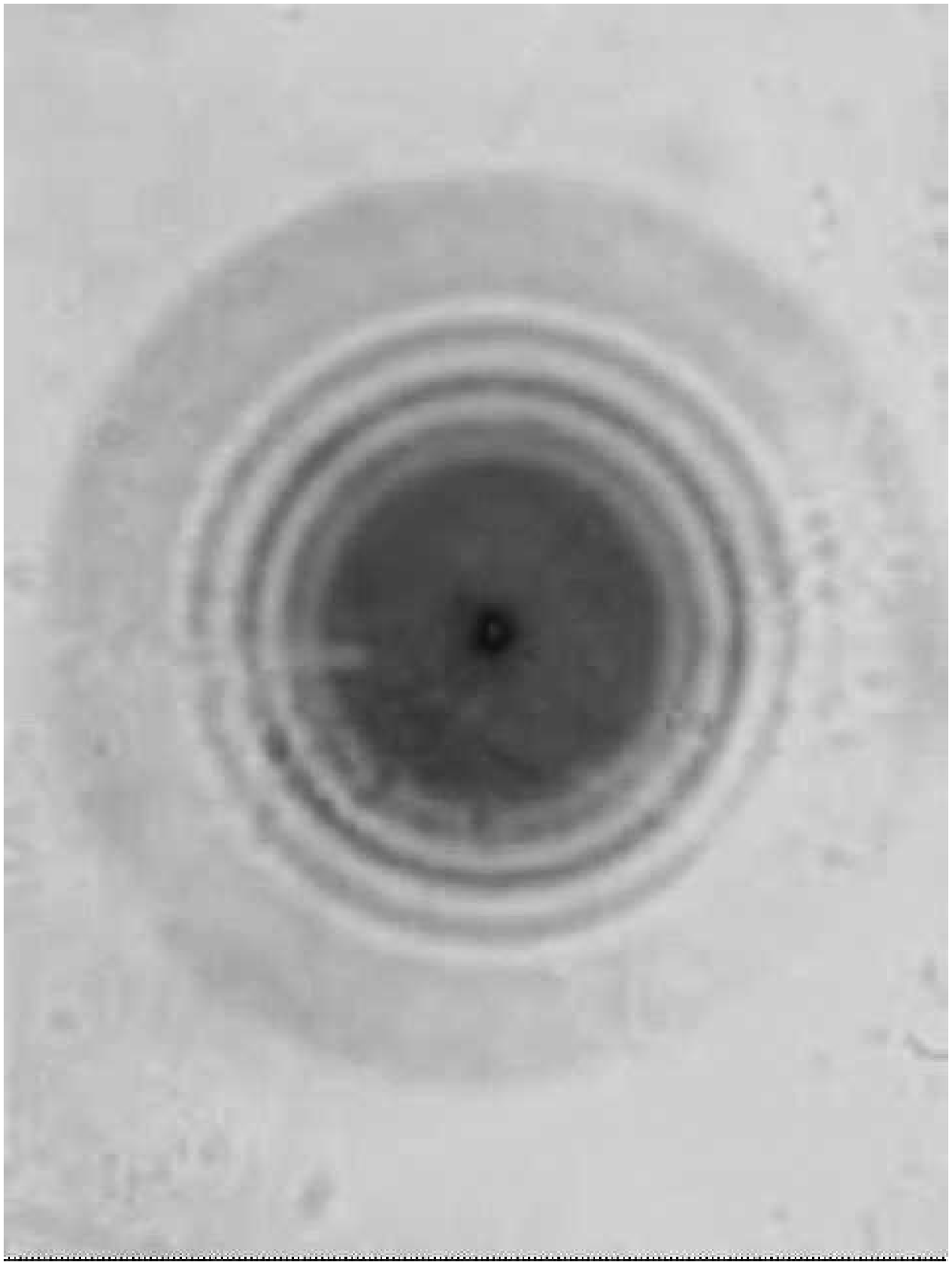} (a)
\epsfig{width = 70mm, file=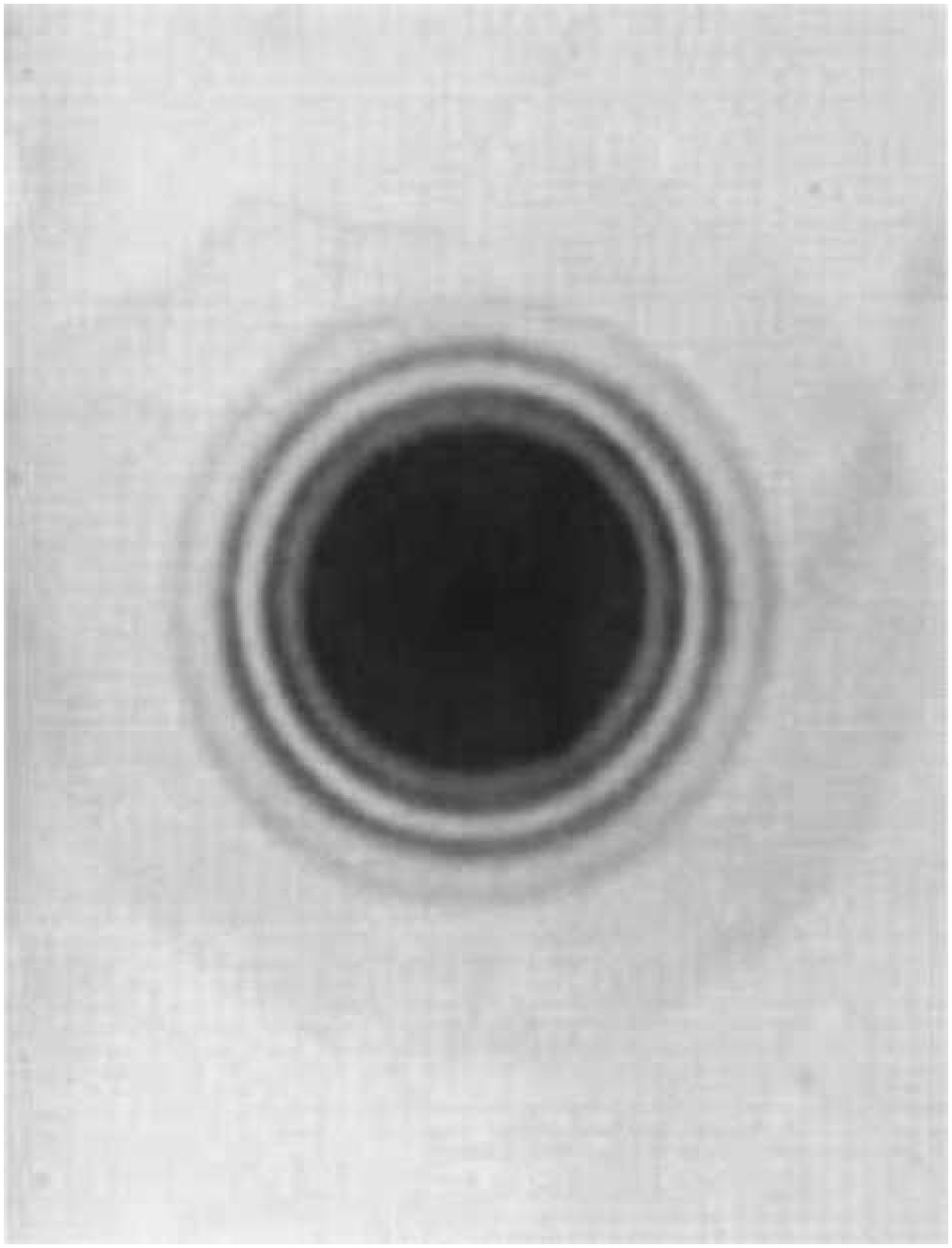} (b)\\
\end{center}
{\bf Fig. 1:} Examples of uranium radiohalos [2]. \textbf{(a)} A fully-developed
uranium radiohalo (dark mica). A uranium radiohalo comprises eight rings,
but some rings are of similar size and cannot easily be distinguished. \textbf{(b)} A
uranium radiohalos in which all four isotopes are from the $^{238}$U decay
series.

\begin{center}
\epsfig{width = 70mm, file=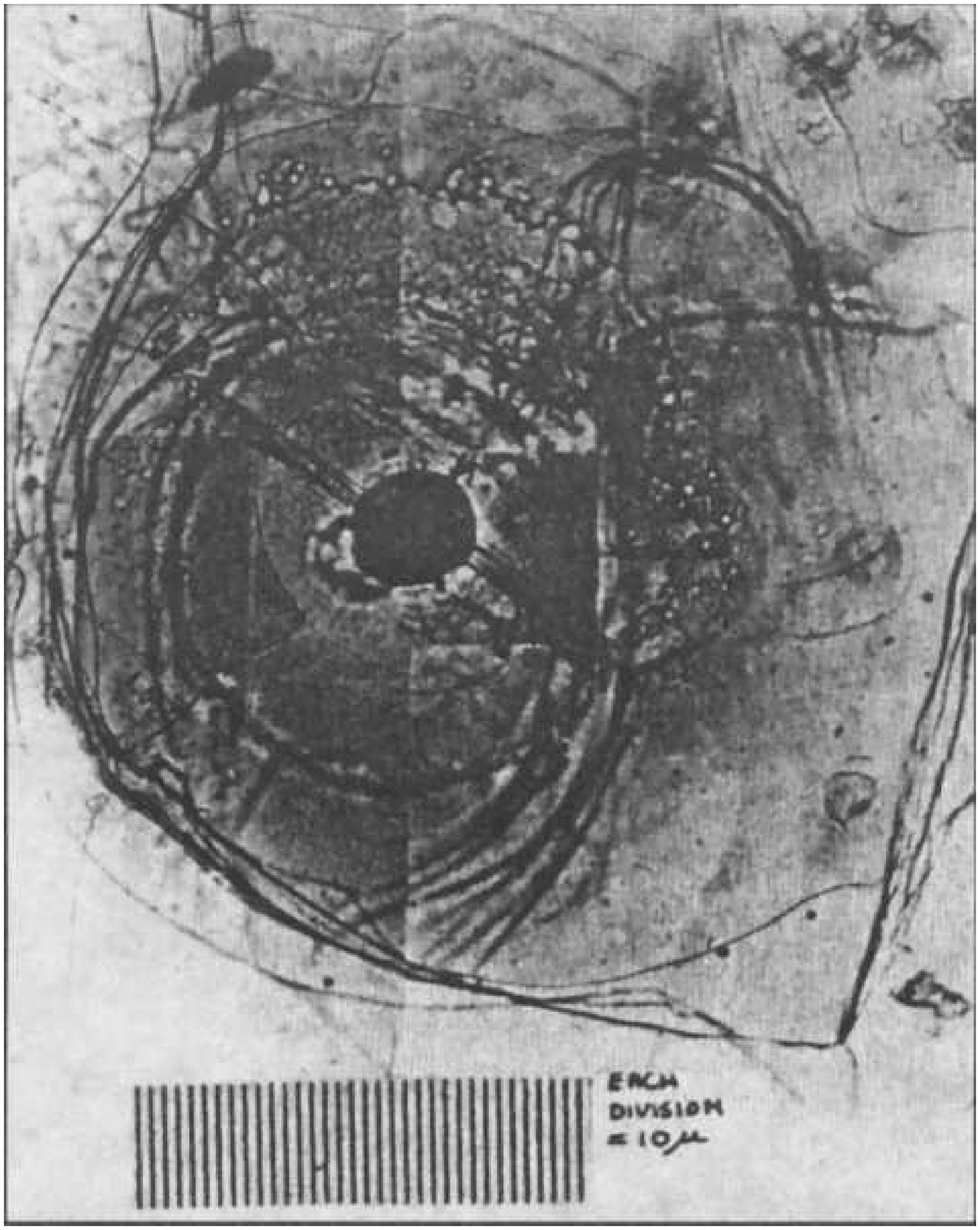} (a)
\epsfig{width = 70mm, file=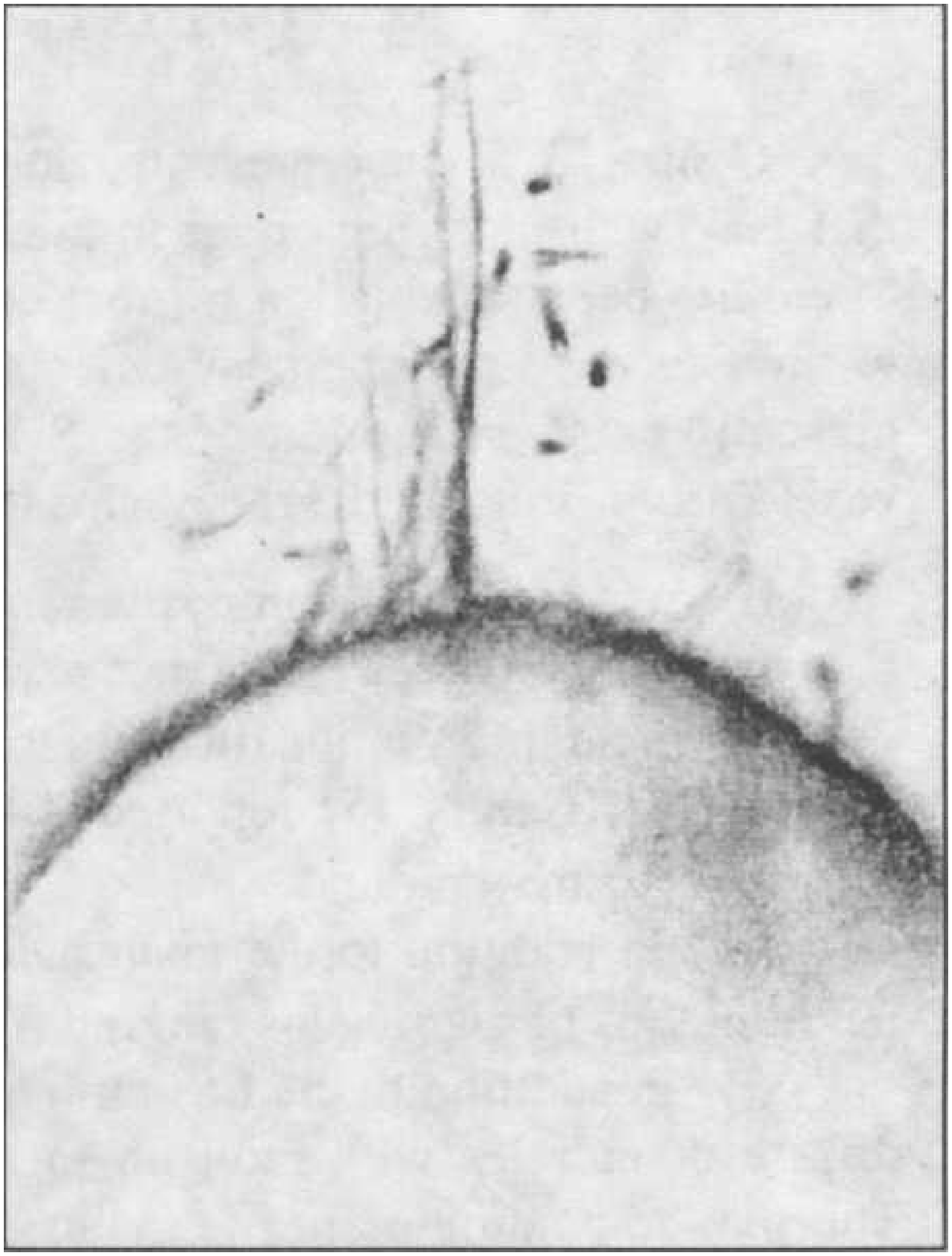} (b)\\
\end{center}
{\bf Fig. 2:}\textbf{(a)}  A photomosaic of a part of one supergiant halo (SGH) F-12
discovered by Grady and Walker [3]. The essential characteristic features of
this SGH are well reproduced by those of PIRH(-) supergiant halos (see the
text); {\bf (b)} A photo image of the mica surface showing fossil tracks next to inclusion pit.

\end{document}